\documentclass{article}
\usepackage[dvips]{graphicx}
\usepackage{amssymb}

\title{Parallel and Perpendicular Lamellae on Corrugated Surfaces}
\author{Yoav Tsori\\
Laboratoire Mati\`ere Molle \& Chimie (UMR 167)\\
ESPCI, 10 rue Vauquelin, 75231 Paris CEDEX 05, France\\
\and
David Andelman\\
School of Physics and Astronomy\\
Raymond and Beverly Sackler Faculty of Exact Sciences\\
Tel Aviv University, Tel Aviv 69978, Israel}

\date{15/4/2003} 

\begin{document}

\setlength{\baselineskip}{14pt} 

\maketitle

\begin{abstract}
We consider the relative stability of parallel and perpendicular
lamellar layers on corrugated surfaces. The model can be applied
to smectic phases of liquid crystals,  to lamellar phases of
short-chain amphiphiles and to lamellar phases of long-chain block
copolymers. The corrugated surface is modelled by having a single
$q$-mode lateral corrugation of a certain height. The lamellae
deform close to the surface as a result of chemical interaction
with it. The competition between the energetic cost of elastic
deformations and the gain in surface energy determines whether
parallel or perpendicular lamellar orientation (with respect to
the surface) is preferred. Our main results are summarized in two
phase diagrams, each exhibiting a transition line from the
parallel to perpendicular orientations. The phase diagrams depend
on the three system parameters: the lamellar natural periodicity,
and the periodicity and amplitude of surface corrugations. For a
fixed lamellar periodicity (or polymer chain length), the parallel
orientation is preferred as the amplitude of surface corrugation
decreases and/or its periodicity increases. Namely, for surfaces
having small corrugations centered at long wavelengths. For a
fixed corrugation periodicity, the parallel orientation is
preferred for small corrugation amplitude and/or large lamellae
periodicity. Our results are in agreement with recent experimental
results carried out on thin block copolymer films of PS-PMMA
(polystyrene-polymethylmethacrylate) in the lamellar phase, and in
contact with several corrugated surfaces.
\end{abstract}

\newpage
\section{Introduction}

Numerous ways to control the orientation of ordered meso-phases
have been extensively studied in recent years. As an example, we
mention that lamellar block copolymers (BCP) confined between two
flat and parallel surfaces have been shown to orient parallel or
perpendicular to the surfaces, depending on the surface separation
\cite{russell1,turner1}. Furthermore, chemically patterned surface
can induce perpendicular or even tilted lamellae, if the chemical
interaction is strong enough \cite{muthu1,TA1}. Applying shear is
yet another effective method in producing large well-aligned
samples \cite{shear}, but this method is difficult to implement in
thin films. Another related situation can be found in
liquid-crystals, where surface anchoring determines the direction
of adjacent molecules and affects the bulk orientation and
possible defects \cite{PGGP}. This type of surface effect has been
 studied extensively in relation to the Fr\'{e}edericksz transition
and in twisted nematic liquid-crystal displays. Electric field can
be very useful in aligning samples in which a large dielectric or
conductivity contrast exists between the components
\cite{AH93,AH94,krausch1,russell2,PW1,TA02,TTAL,TTL,russell3}.
However, the use of electric field requires rather sophisticated
experimental setups in order to avoid adverse effects of ion
accumulation at electrodes, short-circuits due to dust particles,
etc. \cite{russell3}.

Most of the studies mentioned above are restricted to films in
contact with flat and smooth solid surfaces. Little attention has
been paid to the role of surface roughness on film morphology and
orientation, although it is clear that non-flat surface topography
gives rise to defects (e.g. vacancies in the lamellar ordering)
and affects the lamellae orientation. Quite recently, experiments
carried out by Hashimoto and co-workers on lamellar BCP films
\cite{hashimoto} addressed the question of how surface roughness
affects BCP film orientation. In their study it was shown that the
degree of surface roughness controls the lamellar orientation,
leading to situations where the BCP films  orient themselves
parallel or perpendicular to the surface. These experiments serve
as a starting point of our theoretical investigation, where we
restrict ourselves to lamellar multilayer systems in contact with
one corrugated surface. The model is expressed in terms of the
elastic energy of lamellar or smectic systems, and is described by
only two elastic constants. Hence, although some system-specific
details are missing, the results are not restricted to BCP films,
but are more general and equally apply to a broad class of systems
ranging from smectic liquid-crystals to lyotropic
(oil/water/amphiphile) systems. The main difference between these
systems is in the values of the system parameters: the lamellar
periodicity, strength of surface interaction and elastic
constants.

The elastic deformation energy of lamellar layers is
studied separately for parallel and perpendicular
orientations. It is shown that depending on the surface
corrugation amplitude and periodicity, phase transitions
can occur between the two orientations. Since real surfaces
are never ideally flat, understanding and characterizing
surface roughness can be of great importance in controlling
orientation of lamellar phases. This orientation mechanism
is complementary to the mechanisms mentioned above of
electric fields, shear and chemical surface patterning.
In Sections \ref{perp_sec} and \ref{para_sec} we calculate the
deviations of lamellae from their corresponding flat perpendicular
and parallel states.  The free energy of the parallel and
perpendicular states are compared in Sec. \ref{diag_sec}, and a
general discussion and comparison with previous works follows in
Sec. \ref{discussion}.

\section{Perpendicular Layers on Corrugated Surface}\label{perp_sec}

Consider a lamellar system confined by one topographically
corrugated surface, as depicted in Fig.~1. Instead of considering
a rough surface having a random and quenched topography, we assume
that the surface is characterized by a typical corrugation with a
single amplitude and wavelength. The difference between a true
rough surface and a corrugated one is not expected to be very
significant as long as the rough surface amplitude and wavelengths
do not vary much about their average values. A further
simplification is that the surface height $h$ (measured along the
$z$-direction) is taken to depend only on the lateral $x$
direction, while it is translational invariant in the $y$
direction:
\begin{eqnarray}\label{surface}
h(x)=R\cos(q_sx) + h_0 ~,
\end{eqnarray}
\begin{figure}[h!]
\begin{center}
\includegraphics[scale=0.7,bb=65 360 510 600,clip]{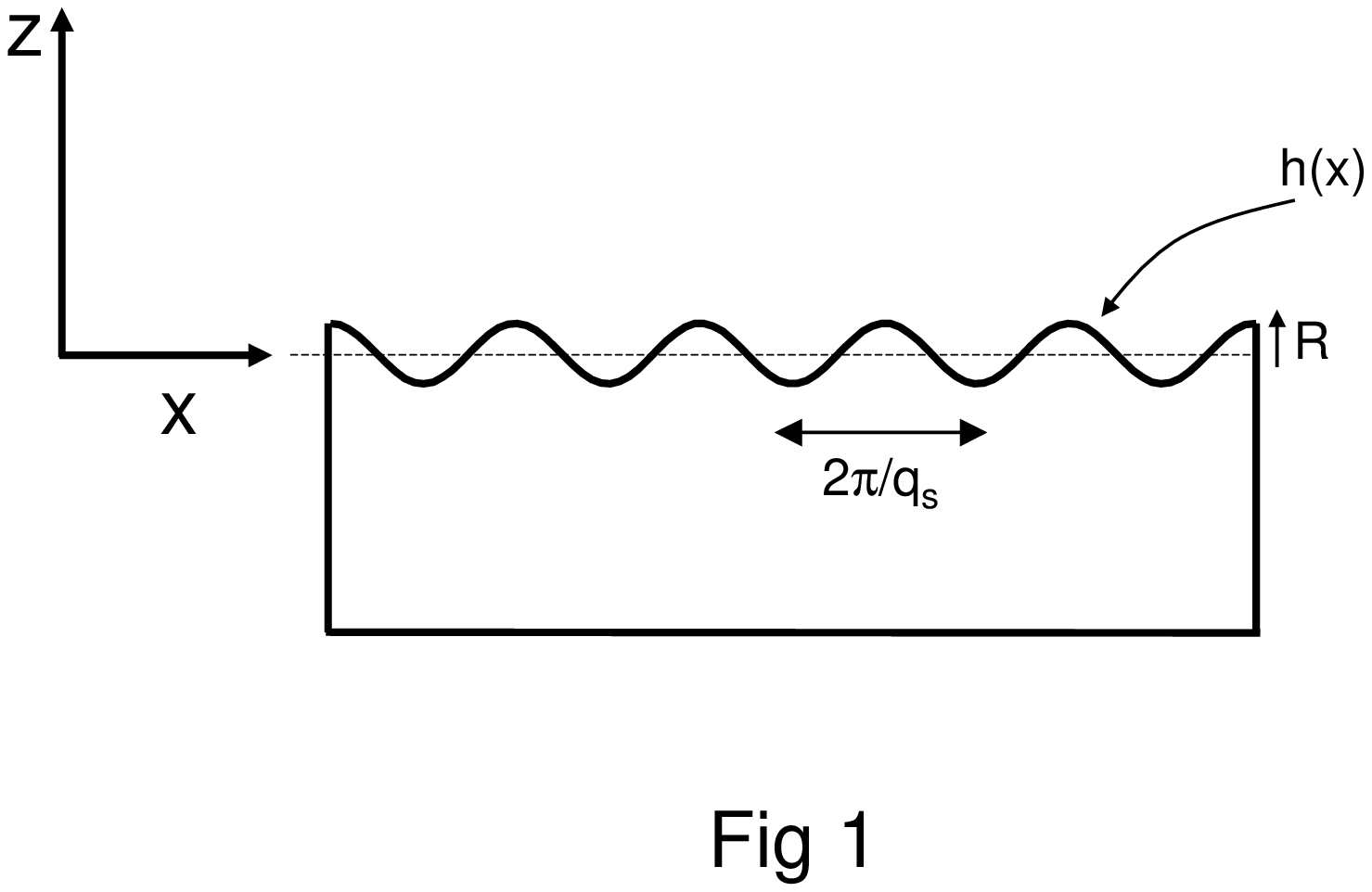}
\end{center}
\caption{\sl Schematic illustration of the rough confining
surface.}
\end{figure}
where the average $h_0=\langle h(x)\rangle$ is taken hereafter to
be zero, $h_0=0$, $R$ is the corrugation amplitude and $d_s\equiv
2\pi/q_s$ is the surface lateral periodicity.

So far we described the quenched corrugated surface geometry. Next
we consider the energetics and structure of a lamellar phase in
contact with such a surface. We start by examining the
order-parameter corresponding to lamellae which are perpendicular
to the average surface position:
\begin{eqnarray}\label{phi_perp}
\phi_\perp({\bf r})=\phi_{0}\cos(q_0x+q_0u(x,z)) ~,
\end{eqnarray}
where $\phi_0$ is the amplitude of lamellar concentration
variations, $d_0\equiv 2\pi/q_0$ is the bulk lamellar
spacing, and $u(x,z)$ is a slowly-varying phase describing
the deviation from flat lamellae, perpendicular to the
surface and given by $\phi_0\cos(q_0 x)$.

The total free--energy $F$ of the lamellar stack in contact with
one surface can be written as a sum of two terms, $F=F_b+F_s$,
where $F_s$ is the surface energy and $F_b$ is the bulk lamellar
contribution. In a lamellar system there are different energy
costs associated with bending and compression of the layers. It
can be shown that for a slowly varying phase $u(x,z)$, an
expansion of the free energy $F_b$ to quadratic order in $u$ and
its spatial derivatives, can be written in complete analogy with
smectic phases of liquid crystals as:
\begin{eqnarray}\label{Fb_perp}
F_b=\frac12\int\left[K\left(u_{zz}\right)^2+B\left(u_x\right)^2\right]
{\rm d}V~.
\end{eqnarray}
In the above equation the integral is over the entire volume, $B$
is the compression modulus and $K$ is the bending modulus. In BCP
systems the model holds well in intermediate and strong
segregations, where the two copolymer blocks are well separated in
different domains and the chains are highly stretched. The elastic
moudlii are $K\sim d_0\gamma_{\rm AB}$ and $B\sim \gamma_{\rm
AB}/d_0$, where $\gamma_{\rm AB}$ is the interfacial tension
between the A and B polymer blocks, and $d_0$ is the lamellar
periodicity \cite{TJ1}. For typical di-BCP such as
polystyrene/polymethylmethacrylate (PS/PMMA), where $\gamma_{\rm
AB}\simeq 2$ mN/m and $d_0\simeq 50$ nm, we estimate their values
to be $K\simeq 10^{-11}$ J/m and $B\simeq 4\cdot 10^5$ J/m$^3$.
For liquid crystals, $K\simeq 2\cdot 10^{-11}$ J/m and $B\simeq
10^7$ J/m$^3$ \cite{collin}, whereas for lyotropic
(water/surfactant/oil) systems, $K$ is about $1~k_BT/d_0\simeq
3\cdot 10^{-30}$ J/m and $B\simeq 1$ J/m$^3$ \cite{ligoure}.
Clearly these lyotropic phases are very `soft' and their elastic
modulii have small values.

The penetration length arising from Eq.~(\ref{Fb_perp}) is
\begin{eqnarray}\label{lambda}
\lambda\equiv (K/B)^{1/2}~.
\end{eqnarray}
In lamellar di-BCP this length is proportional to the lamellar
spacing, $\lambda\sim d_0\simeq 50$ nm. In lyotropic systems this
distance is very small because of the reduced rigidity.

The second term in the free energy is the surface contribution.
Similar to previous works \cite{TA1,PW1,TA02}, we assume a
short--range surface field coupled linearly with the lamellar
order--parameter at the surface.

\begin{eqnarray}\label{Fs}
F_s=\int \sigma\phi~ {\rm d}S ~.
\end{eqnarray}
The parameter $\sigma$ is the surface field and the integral is
taken over the entire corrugated surface.  The value of $\sigma$
is taken to be a constant throughout the surface, describing a
corrugated surface which is chemically homogeneous. A positive
(negative) $\sigma$ favors adsorption of negative (positive)
$\phi$ at the surface. In the language of an A/B di-BCP, this
means that B (or A) monomers are preferentially adsorbed on the
surface for $\sigma>0$ ($\sigma<0$). Assuming small distortions in
the order parameter close to the surface, $F_s$ can be expanded to
first order in $u$
\begin{eqnarray}\label{Fs2}
F_s\simeq\sigma\phi_0\int \left[\cos(q_0x)-q_0u\sin(q_0x)\right]~
{\rm d}S
\end{eqnarray}

The entire profile $\phi(x,z)$ can now be calculated by using a
variational principle on the bulk free--energy,
Eq.~(\ref{Fb_perp}). In terms of the phase $u(x,z)$, the resulting
Euler-Lagrange differential equation away from the corrugated
surface is,
\begin{eqnarray}\label{gov_perp}
\lambda^2u_{zzzz}-u_{xx}=0
\end{eqnarray}
This variation has to be complemented by a set of rather complex
boundary conditions. They are obtained by taking the variation of
the full $F_b+F_s$ on the corrugated surface, $z=h(x)$, defined in
Eq.~(\ref{surface}),
\begin{eqnarray}\label{perp_bc1}
\frac{\partial f_b}{\partial u_{zz}}&=&0\\
\frac{\partial f_s}{\partial u}+\left(\frac{\partial
f_b}{\partial\nabla u}-\frac{\partial}{\partial z} \frac{\partial
f_b}{\partial u_{zz}}~\hat{z}\right)\cdot
\hat{n}&=&0\label{perp_bc2}
\end{eqnarray}
The unit vector normal to the surface is defined as:
$\hat{n}=-(q_sR\sin q_sx,1)/\sqrt{(q_sR\sin q_sx)^2+1}$, and $f_b$
and $f_s$ are the integrand of the volume and surface integrals,
Eqs.~(\ref{Fb_perp}) and (\ref{Fs2}), respectively.

In order to proceed we need to make some further simplifications.
Solving the partial differential equation, Eq. (\ref{gov_perp}),
with the complex boundary condition,
Eqs.~(\ref{perp_bc1})-(\ref{perp_bc2}) is an extremely difficult
task. A further simplification is to assume that $u(x,z)$ is given
by the single bulk mode $q_0$ in the $x$ direction, while it
contains a sum over all possible Fourier modes in the $z$
direction:
\begin{eqnarray}\label{perp_ansatz}
u(x,z)={\rm e}^{iq_0 x}\sum_k A_k {\rm e}^{ikz}~~+~~c.~c.
\end{eqnarray}
This simple sinusoidal form of $u$ along the surface $x$ direction
may not fully account for the incommensurability that exists
between the lamellar and surface periodicities. Defects and
vacancies in the lamellar ordering along the surface are not
included either. Equations (\ref{gov_perp}) and
(\ref{perp_ansatz}) lead to a selection of specific $k$ modes
depending on the bulk mode $q_0$ and the penetration length
$\lambda$:
\begin{eqnarray}
k^4+\lambda^{-2}q_0^2=0
\end{eqnarray}
This equation has two different roots, $k_\pm$ (and their two
complex conjugates)
\begin{eqnarray}
k_\pm&=&\pm\left(\frac{q_0}{\lambda}\right)^{1/2}\cdot {\rm e}^{\pm
i\pi/4}
\end{eqnarray}
where $\exp(i\pi/4)=(1+i)/\sqrt{2}$~ is the 4th root of
unity in the complex plane. The first boundary condition,
Eq. (\ref{perp_bc1}), implies that $A_{k_{+}}=A_{k_{-}}$.
Throughout this paper we assume $q_sR<q_0R\ll 1$ , whence
it is valid to approximate $\sqrt{(q_sR\sin
q_sx)^2+1}\simeq 1$ and $\exp({ikR \cos(q_sx)})\simeq 1$.
The second boundary condition [Eq.(\ref{perp_bc2})] then
gives
\begin{eqnarray}
-\sigma\phi_0 q_0\sin q_0x-iBq_sR\sin q_sx\cdot\sum_k q_0 A_k {\rm
e}^{i(q_0x+kz)}-iK\cdot \sum_k k^3 A_k {\rm e}^{i(q_0x+kz)}&=&0
\end{eqnarray}
The $z$-dependent term in the exponentials is neglected as
can be justified for $q_0R\ll 1$. The second term
(proportional to $B$) can also be neglected for $q_sR\ll
1$, leading to the final expression of the lamellar order
parameter:
\begin{figure}[h!]
\begin{center}
\includegraphics[scale=0.5,bb=130 95 505 775,clip]{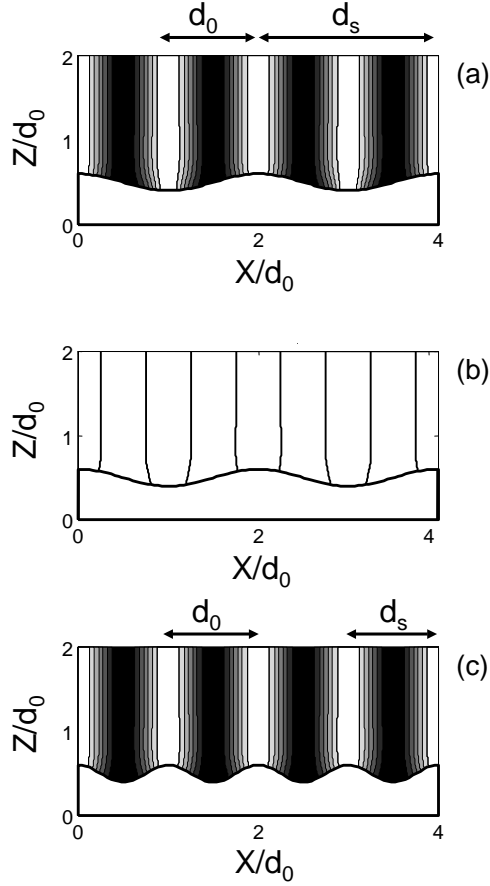}
\end{center}
\caption{\sl Perpendicular lamellar layers in contact with a
sinusoidal surface, as obtained from Eqs. (\ref{phi_perp}) and
(\ref{u_perp}). All length are scaled by $d_0$, the lamellar bulk
periodicity, and the origin of the $z$ axis is shifted by $d_0/2$
for clarity purposes. (a) The lamellar periodicity $d_0$ is chosen
to be half of the surface one, $d_0=d_s/2$. Different grey shades
correspond to different values of the order parameter $\phi$ of
Eq. (\ref{phi_perp}), where dark and white correspond to negative
and positive $\phi$, respectively. (b) Inter-material dividing
surface corresponding to lines where $\phi=0$ in part (a). (c) The
same as in (a), but the lamellar periodicity equals the surface
one, $d_s=d_0$. The parameters used are: $B=4$, $K=B/(4q_0^2)$ and
$\sigma=\sqrt{BK}/4$. }
\end{figure}
\begin{eqnarray}\label{u_perp}
u&=&A_0{\rm e}^{i(q_0x+k{_+}z)}+A_0{\rm e}^{i(q_0x+k{_-}z)}~~+~~c.~c.\\
A_0&=&\frac{\sigma\phi_0 q_0}{2K(k_{+}^3+k_{-}^3)}\nonumber
\end{eqnarray}
where  $A_{q_0}$ is denoted  $A_0$ in order to simplify the
notation. The full order-parameter expression is obtained by
substituting $u$ from above in Eq.~(\ref{phi_perp}). The decay
length of lamellar undulations is proportional to the lamellar
spacing, $1/k_\pm \sim (\lambda/q_0)^{1/2}\sim d_0$.

The perpendicular lamellar stack on a rough surface is
shown in Fig.~2. In Fig.~2(a) the lamellar periodicity is
half of the surface one. The surface is attractive to the
component marked in dark shades, causing them to expand in
its vicinity. Light lamellae are in turn contracted close
to the surface; this behavior is seen in part 2(c) too,
where  the surface periodicity matches the bulk lamellar
one. Clearly,  the curvature of the lamellae adjusts to
the surface one in order to achieve the best compromise
between elastic deformation and surface coverage.

\section{Parallel Layers on Corrugated Surface}\label{para_sec}

We now turn to describe a parallel lamellar stack in contact with
the same sinusoidally corrugated  surface. Our derivation is
related to previous treatments \cite{PGGP,TJ2,turner3}, but the
boundary conditions are handled differently by introducing the
same type of surface term in the free energy $F_s$, as in the
previous section. The lamellae can be described along the same
lines as for the perpendicular case but keeping in mind the
important points where the stack orientation affects the free
energy:
\begin{eqnarray}
\phi_\parallel({\bf r})=-\phi_0\cos(q_0z+q_0u(x,z))
\end{eqnarray}
The perfect parallel layers, $\phi_\parallel({\bf r})=-\phi_q\cos
q_0z$ are recovered far from the surface (where $u=0$). The bulk
lamellar phase free-energy is obtained simply by interchanging the
roles of $x$ and $z$ axes in the free energy of Eq.~(3): $x
\leftrightarrow z$
\begin{eqnarray}\label{Fb_para}
F_b=\frac12\int\left[K\left(u_{xx}\right)^2+B\left(u_z\right)^2\right]
{\rm d}V
\end{eqnarray}
As  for the case of perpendicular lamellae, the surface energy,
Eq.~(\ref{Fs}), can be expanded in small $u$ as follows
\begin{eqnarray}
F_s\simeq \sigma\phi_0\int \left[q_0u\sin(q_0z)-\cos(q_0z)\right]~
{\rm d}S
\end{eqnarray}
The governing equation for $u$ is obtained from a variation
principle applied to Eq.~(\ref{Fb_para}), $\delta F_b/\delta u=0$
\begin{eqnarray}\label{gov_para}
\lambda^2u_{xxxx}-u_{zz}&=&0
\end{eqnarray}
with the boundary condition obtained from a variation of $F_b+F_s$
on the $z=h(x)$ corrugated surface.

\begin{eqnarray}\label{para_bc}
\frac{\partial f_s}{\partial u}+\left(\frac{\partial
f}{\partial\nabla u}-\frac{\partial}{\partial x} \frac{\partial
f}{\partial u_{xx}}~\hat{x}\right)\cdot \hat{n}&=&0
\end{eqnarray}

Writing $u$ as $u(x,z)=A_0{\rm e}^{i\alpha z}\cos q_sx$ and using
Eq. (\ref{gov_para}) we find that $\alpha=i\lambda q_s^2$. At this
point it should be emphasized that when the surface periodicity is
larger than the lamellar periodicity ($q_s<q_0$), the decay length
$1/\alpha\sim q_0/q_s^2$ for the parallel stack is much larger
than the decay length $1/k_\pm\sim 1/q_0$ in the perpendicular
case. This, in turn, means that for fixed $R$ the elastic
deformation gives preference to perpendicular ordering.

As in the case of perpendicular lamellae, we use $q_sR<q_0R\ll 1$
to approximate \\
$\sqrt{(q_sR\sin q_sx)^2+1}\simeq 1$ and obtain from the
boundary condition, Eq. (\ref{para_bc}),
\begin{eqnarray}
\sigma\phi_0 q_0^2R + B\lambda q_s^2 A_0 =0
\end{eqnarray}
yielding the order-parameter of parallel lamellae
\begin{eqnarray}\label{u_para}
u(x,z)&=&A_0{\rm e}^{i\alpha z}\cos q_sx\\
A_0&=&-\frac{\sigma\phi_0 q_0^2 R}{B\lambda
q_s^2}=-R\frac{\sigma\phi_0}{\sqrt{BK}}\left(\frac{q_0}{q_s}\right)^2
\nonumber
\end{eqnarray}

The parallel layering given by Eq. (\ref{u_para}) is plotted in
Fig.~3. In 3(a) [as in Fig.~2(a)] the surface periodicity is twice
larger than the lamellar one, and in 3(b) the two periodicities
are equal. In the former case, dark regions (negative $\phi$)
appear close to the wall, the lamellae are able to closely follow
the surface contour and distortions are long-range. In the latter
case the surface topography changes too quickly for the lamellae
to follow and the lamellae lie almost perfectly flat. Distortions
of the stack, in this case, can be seen only in the very close
vicinity of the surface.

\begin{figure}[h!]
\begin{center}
\includegraphics[scale=0.8,bb=130 185 400 580,clip]{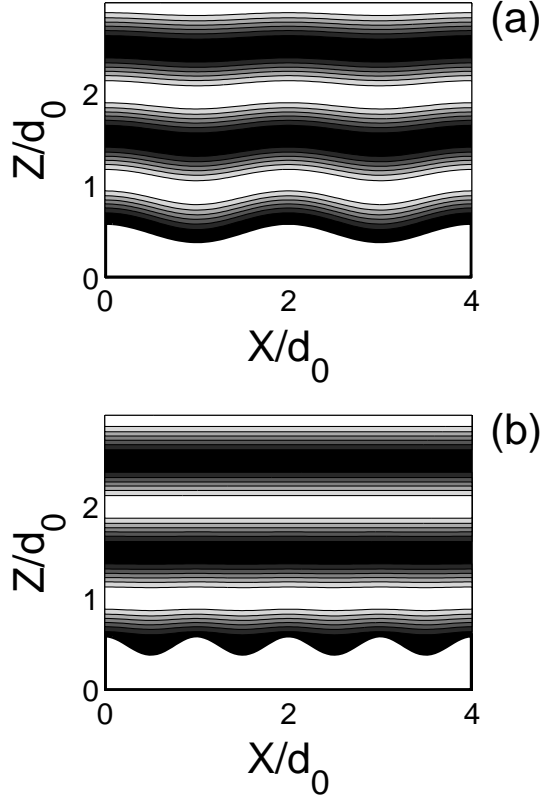}
\end{center}
\caption{\sl Parallel lamellar layers in contact with a corrugated
surface, calculated from the free energy minimization of Sec. 3.
(a) The lamellar periodicity is half of the surface one
$d_0=d_s/2$, and the lamellae follow the surface topography. (b)
The two periodicities are equal, $d_s=d_0$, and the lamellae are
unable to follow the surface height variations. Parameters here
are the same as in Fig.~2.  }
\end{figure}
%

\section{Phase Diagram}\label{diag_sec}

The elastic energy of the lamellar phase in the two orientations
is obtained by substitution of Eq. (\ref{u_perp}) in Eqs.
(\ref{Fb_perp}) and (\ref{Fs}) (perpendicular lamellae) and Eq.
(\ref{u_para}) in Eqs. (\ref{Fb_para}) and (\ref{Fs}) (parallel
lamellae). Depending on the system parameters $R$, $q_s$, $q_0$
and $\sigma$, the minimum of the free energy is obtained for
either one of the two orientations.

Before we proceed in presenting the corresponding phase diagram,
we note that our derivation is not valid over the entire $R$,
$q_s$, $q_0$ and $\sigma$ parameter space. The assumption of small
distortions, $q_0u\ll 1$, together with Eq.~(\ref{u_para}) implies
that ${R\sigma\phi_0}q_0^3q_s^{-2}\ll \sqrt{BK}$. This can be
rewritten as $q_s R\gg (q_0R)^{3/2}
(\sigma\phi_0/\sqrt{BK})^{1/2}$. Combining these two inequalities,
we obtain that the limits of validity of our derivation are given
by
\begin{eqnarray}
1\gg
q_0R>q_sR\gg(q_0R)^{3/2}\left(\frac{\sigma\phi_0}{\sqrt{BK}}
\right)^{1/2}
\end{eqnarray}
Cast in different terms this can be written as
\begin{eqnarray}\label{validity}
\left(\frac{\sigma\phi_{0}}{\sqrt{BK}}\right)^{1/2}(q_0R)^{1/2}
\ll \frac{q_s}{q_0}<1
\end{eqnarray}
\begin{figure}[h!]
\begin{center}
\includegraphics[scale=0.5,bb=85 250 500 670,clip]{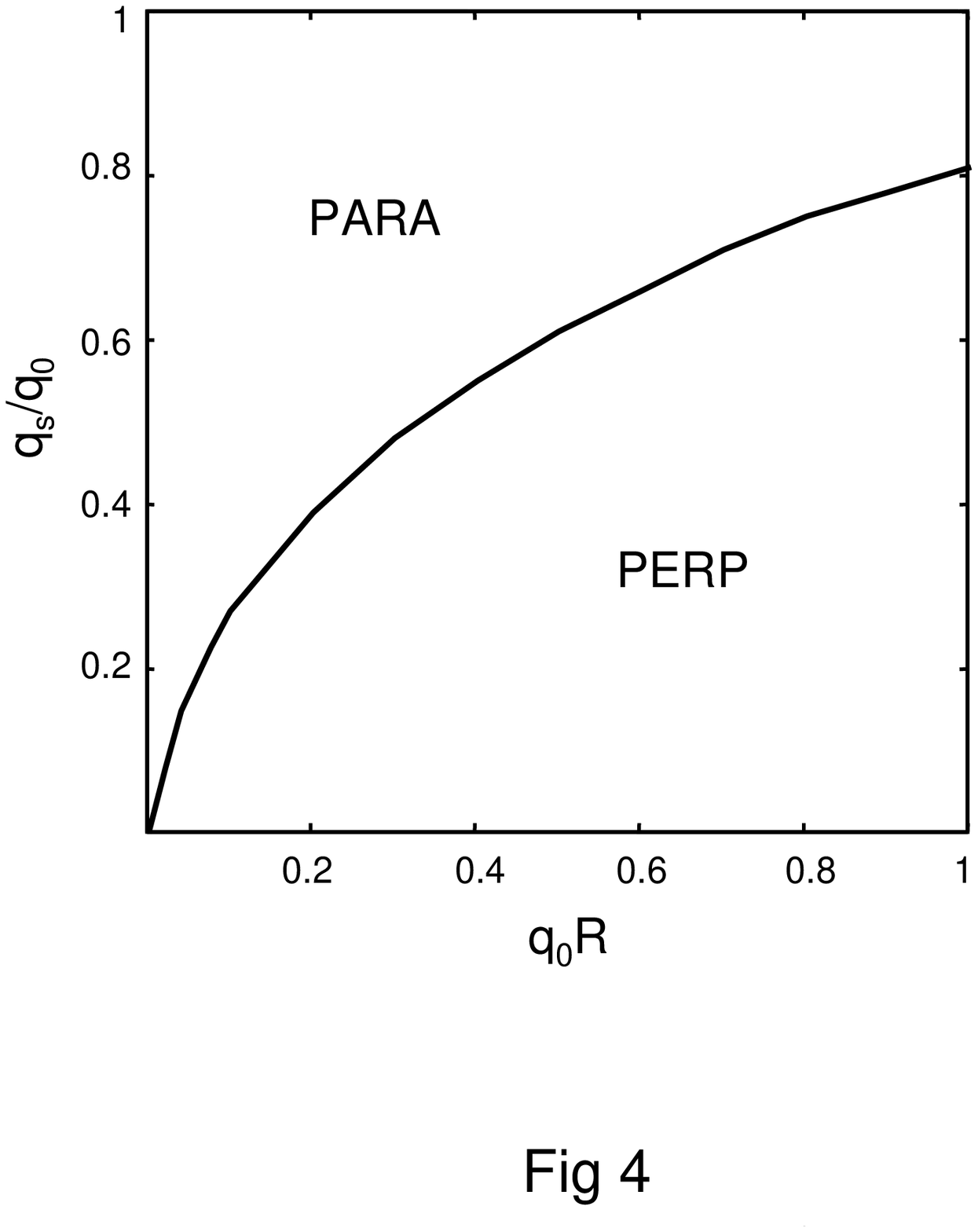}
\end{center}
\caption{\sl Phase diagram in terms of  $q_s$ and $R$ with
fixed surface field $\sigma$ and lamellar periodicity
$2\pi/q_0$. For fixed value of $q_s$ (scaled by $q_0$), a
horizontal scan of increasing $R$ (scaled by $1/q_0$) will
increase the elastic penalty of parallel layers, and
favors perpendicular ordering. On the other hand, a
vertical scan of increasing $q_s$ (while keeping $R$
constant) limits the deformations in the parallel state to
the surface, and favors parallel ordering.
$\sigma=2\sqrt{BK}$ and all other parameters are the same
as in Figs. 2 and 3. }
\end{figure}

Before presenting the calculated phase diagram, it is of use to
present a rough estimation of the parallel and perpendicular
free--energies. Disregarding numerical prefactors, the bulk free
energy of the perpendicular state is
\begin{eqnarray}
\frac{F_\perp}{S}&\sim&A_0^2\left(Bq_0^2+Kk_\pm^4\right)\cdot\frac{1}{{\rm
Im}( k_
\pm)}\nonumber\\
&=&\frac{\sigma^2\phi_0^2}{K}\frac{1}{q_0}
\end{eqnarray}
where $S$ is the surface area and $A_0$ is taken from Eq.
(\ref{u_perp}). At  distances from the surface greater
than $1/{\rm Im}(k_\pm)$, the distortion $u$ is
negligible, and the factor $1/{\rm Im}(k_\pm)$ represents
the effective volume to area ratio of the integration.

The free energy of the parallel state is similarly estimated to be
\begin{eqnarray}
\frac{F_\parallel}{S}&\sim&A_0^2\left(B\alpha^2+Kq_s^4\right)\cdot\frac{1}
{{\rm Im}(\alpha)}\nonumber\\
&=&\frac{\sigma^2\phi_0^2}{K}\frac{1}{q_0}\left(\frac{q_0}{q_s}
\right)^2\left(q_0R\right)^2
\end{eqnarray}
Here $A_0$ is taken from Eq. (\ref{u_para}) and the effective
volume to area ratio is $1/{\rm Im}(\alpha)$.

The difference between the two free energies is roughly
proportional to
\begin{equation}\label{scaling}
F_\parallel-F_\perp~\propto~ \left(\frac{q_0}{q_s}
\right)^2\left(q_0R\right)^2-1
\end{equation}
Eq.~(\ref{scaling}) describes qualitatively the system behavior.
If $q_0$ is fixed while $R$ and $q_s$ can vary, it directly
follows that for $q_0R>q_s/q_0$ the perpendicular state is
favored, while for $q_0R<q_s/q_0$ the parallel state is favored.
Similar relations hold when $q_s$ is fixed and $R$ and
$q_0$ are allowed to vary, or when $R$ is fixed but $q_s$
and $q_0$ can vary.
%
%
%
%

The above simplified approach describes qualitatively the system
behavior and may be used as a `rule of thumb'. Furthermore, the
surface energies and the correct numerical prefactors can be taken
into account as well. Figure~4 shows the phase diagram in the
($q_s,R$) plane, for fixed $q_0$ and $\sigma$. Not the entire
shown phase diagram is within the range of validity discussed
above, Eq. (\ref{validity}). For small values of $q_s$, the
perpendicular layering is favored, because of the long-ranged
elastic strain  pertaining in parallel lamellae as compared to
perpendicular ones. As $q_s$ increases, the strain in the parallel
state becomes more restricted to the vicinity of the surface,
until, eventually, the parallel state becomes more stable. At this
transition point, the energetic gain of having a commensurate
layer close to the surface overcomes the loss of elastic energy
deformation.

Different conclusion can be drawn for small values of surface
amplitude $R$, which generally induce a parallel state. Keeping
$q_s$ fixed and gradually increasing $R$ means that the elastic
energy of deforming parallel lamellae increases, while the surface
interaction stays constant. Therefore, at a certain threshold
value of $R$ there is a transition from parallel to perpendicular
ordering. For larger $q_s$ values this critical $R$ value
increases as well.
\begin{figure}[h!]
\begin{center}
\includegraphics[scale=0.5,bb=90 265 490 690,clip]{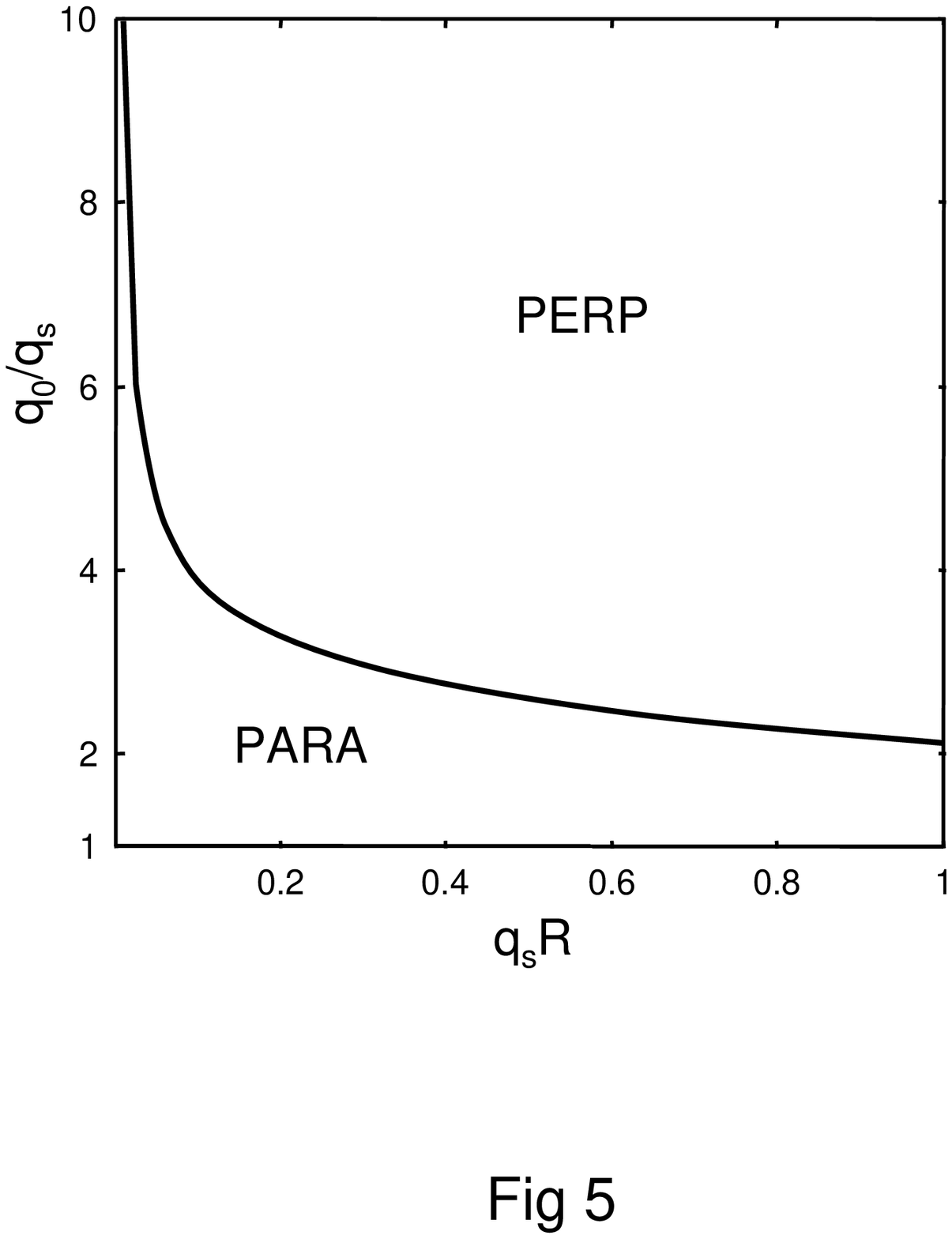}
\end{center}
\caption{\sl Phase-diagram in the $(R,q_0)$ plane. The
surface field $\sigma$ and surface periodicity $2\pi/q_s$
are kept constant. Horizontal scans of increasing $R$
(scaled by $1/q_s$) while keeping $q_0$ fixed, increases
the elastic penalty of parallel layers and favors
perpendicular ordering. Same is true for vertical scans of
increasing $q_0$ (scaled by $q_s$) with fixed $R$.
$\sigma=0.2\sqrt{BK}$ and other parameters are the same as
in Figs. 2 and 3. }
\end{figure}

Figure 5 is a phase diagram in the $(R, q_0)$ plane, with
$q_s$ kept fixed. For a given BCP chain length (fixed
$q_0$), increase of $R$ will also increase the elastic
energy of deforming parallel lamellae, and promotes
perpendicular layering. On the other hand, keeping $R$
fixed and decreasing $q_0$ (so that it becomes comparable
to $q_s$) decreases the range of parallel deformation;
thus, favoring the parallel state.

\section{Discussion}\label{discussion}

Lamellar stacks of either liquid crystals (smectics), short
chains amphiphiles or long-chain BCP undergo deformation
as they try to adjust to the presence of a rough
(corrugated) surface. The amplitude and spatial extent of
inplane and out-of-plane deformations are different and
the preference for parallel or perpendicular orientations
depends on specific bulk and surface parameters. Which
morphology prevails depends on the surface periodicity
$d_s=2\pi/q_s$, surface amplitude $R$, lamellar
periodicity $d_0=2\pi/q_0$, as well as the surface field
strength $\sigma$.

For lamellae oriented perpendicular to the surface we use a model
that indicates how the stack deformation propagates from the
surface into the semi-infinite bulk. Parallel lamellae are studied
using a modification of a model used previously for smectics
phase. In both cases small surface corrugation is assumed such
that we can use the limits: $q_sR\ll 1$, $q_0R\ll 1$. Our study is
complementary to a numerical study by Podariu and Chakrabarti
\cite{chakra1} who studied in detail extremely thin films
(thickness comparable or smaller than the lamellar thickness) of a
lamellar stack.

The system behavior in terms of the parameters $R$ and $q_s$ is
given in the phase diagram of Fig.~4. For fixed surface
periodicity $2\pi/q_s$, increase in the surface amplitude $R$
leads to preference of perpendicular lamellae. On the other hand,
keeping $R$ fixed and increasing $q_s$ leads to preference of
parallel lamellae. This is a consequence of the diminished decay
length of surface-induced undulations. When the undulations of
parallel lamellae are restricted to the surface vicinity, the
energy penalty of the elastic defect can be small. Thus, the phase
transition between these two states is described by a line in the
two-dimensional phase diagram. Quite generally, as the surface
interaction parameter $\sigma$ is increased, this line moves
towards the $R$ axis in such a way that the state of parallel
lamellae occupies a larger region in the phase diagram.

The transition from parallel to perpendicular orientation as
function of $R$ and $q_0$ is given in Fig.~5. Increase of $R$
deforms the parallel layers and generally promotes perpendicular
ordering. On the other hand, increase of $q_0$ while keeping $R$
and $q_s$ fixed, implies a reduced range of parallel deformations
and yield a preference for parallel lamellae.

Rough surface can be used to obtain morphologies that are usually
controlled by chemical means: not only the transition between
parallel and perpendicular lamellae, but also tilted lamellar
morphologies. Indeed, in the limit of very small lamellar
periodicity (large $q_0$) and strong interfacial interactions
$\sigma$, lamellae will appear locally perpendicular to the
surface, and therefore tilt when the surface is not horizontal.
Similar tilted lamellae have been predicted for a BCP melt
confined by chemically patterned surfaces \cite{muthu1,TA1}, as
the system tries to match the lamellae with the stripes.

The present work is motivated by and is of direct relevance to
recent experiments of Hashimoto and co-workers on PS/PMMA
symmetric di-BCP on rough surfaces \cite{hashimoto}. In the
experiments, perpendicular orientation was observed for system
parameters estimated to be $q_s\simeq 0.04$ nm$^{-1}$, $q_0\simeq
0.33$ nm$^{-1}$ and $R\simeq 7.5$ nm. This set of data is in
complete accord with the scaling formula, Eq.~(\ref{scaling}),
because $q_0^4R^2/q_s^2-1$ is large and positive. This also can be
confirmed by the phase diagrams presented in Figs.~4 and 5.

For different surfaces and BCP films reported in Ref.
\cite{hashimoto}, $q_s\simeq 0.018$ nm$^{-1}$, $q_0\simeq 0.19$
nm$^{-1}$ and $R\simeq 2.7$ nm, and the film orients itself in
parallel layers. The scaling formula, Eq.~(\ref{scaling}) yields
here a small positive number, implying weak preference for
perpendicular layers. The phase diagram also show marginal
behavior, possibly preferential for perpendicular layers. This
discrepancy can have several (yet unknown) origins. First, the
exact value of the surface interaction parameter $\sigma$ is not
known from experiments, and it may be different than the values
chosen by us in Figs.~4 and 5. Furthermore, we employed several
approximations in our calculations ignoring, for example, possible
vacancies in the perpendicular lamellae and the temperature
dependence of the elastic modulii. We also assumed that the system
is semi-infinite in the $z$-direction, while in experiments the
BCP film thickness is finite. It may be that the BCP/air free
surface can induce islands or other types of surface-induced
defects that alter the simple picture employed by us. The
free--energy model holds for BCPs in intermediate and strong
segregations, but is less adequate for weaker segregations, as is
possibly the case in the experiments. Lastly, our theory can be
improved by taking into account more realistic surface roughness,
instead of a single surface $q$-mode.

We hope that additional and detailed experiments of lamellar
systems in contact with well characterized rough surfaces will
shed more light on this problem, and will further motivate
theoretical studies of these intriguing systems.

\bigskip
{\it Acknowledgements.~~~~} We thank T. Hashimoto and E. Sivaniah
for many discussions and correspondence, and for sharing with us
their experimental results prior to publication. We thank M.
Cloitre, J.-B. Fournier, F. Tournilhac and L. Leibler for useful
discussions and comments. DA wishes to acknowledge support from
the Israel Science Foundation under grant no. 210/02, and the
Alexander von Humboldt Foundation, while YT thanks the
Chateaubriand fellowship program.


\newpage

\begin{thebibliography}{0}

\bibitem{russell1} Lambooy, P.; Russell, T. P.; Kellogg, G. J.;
Mays, A. M.; Gallagher, P. D.; Satija, S. K. {\it Phys. Rev.
Lett.} {\bf 1994}, {\it 72}, 2899.

\bibitem{turner1} Turner, M. S. {\it Phys. Rev. Lett.} {\bf 1992},
{\it 69}, 1788.

\bibitem{muthu1} Petera, D; Muthukumar, M. {\it J. Chem. Phys.} {\bf
1998}, {\it 109}, 5101.

\bibitem{TA1} Tsori, Y.; Andelman, D. {\it J. Chem. Phys.} {\bf 2001},
{\it 115}, 1970.

\bibitem{shear} Patel, S. S.; Larson, R. G.; Winey, K. I.;
Watanabe, H. {\it Macromolecules} {\bf 1995}, {\it 28}, 4313.
Koppi, K. A.; Tirrell, M.; Bates, F. {\it Phys. Rev. Lett} {\bf
1993}, {\it 70}, 1449. Riise, B. L.; Fredrickson, G. H.; Larson,
R. G.; Pearson, D. S. {\it Macromolecules} {\bf 1995}, {\it 28},
7653.

\bibitem{fred} Fr\'{e}edericksz, V.;  Zolina, V. {\it Trans. Faraday
Soc.} {\bf 1933}, {\it 29}, 919.

\bibitem{PGGP} de Gennes, P. G.; Prost, J. {\it The Physics of
Liquid Crystals}, 2nd Edition (Oxford University Press, Oxford,
1993).

\bibitem{AH93}  Amundson, K.; Helfand, E.; Quan, X; Smith,  S. D. {\it
Macromolecules} {\bf 1993}, {\it 26}, 2698.

\bibitem{AH94} Amundson, K.; Helfand, E.; Quan, X.; Hudson, S. D.;
Smith, S. D. {\it Macromolecules} {\bf 1994}, {\it 27}, 6559.

\bibitem{krausch1} B\"{o}ker, A.; Knoll, A.; Elbs, H.; Abetz, V.;
M\"{u}ller, A. H. E.; Krausch, G. {\it Macromolecules} {\bf 2002},
{\it 35}, 1319.

\bibitem{russell2} Thurn-Albrecht, T.; Schotter, J.; K\"{a}stle, G.
A.; Emley, N.; Shibauchi, T.; Krusin-Elbaum, L.; Guarini, K.;
Black, C. T.; Tuominen, M. T.; Russell, T. P. {\it Science} {\bf
2000}, {\it 290}, 2126.

\bibitem{PW1} Pereira, G. G. ; Williams, D. R. M. {\it Macromolecules}
{\bf 1999}, {\it 32}, 8115. Ashok, B.; Muthukumar, M.; Russell, T.
P. {\it J. Chem. Phys.} {\bf 2001}, {\it 115}, 1559.

\bibitem{TA02} Tsori, Y.; Andelman, D. {\it Macromolecules} {\bf
2002}, {\it 35}, 5161.

\bibitem{TTAL} Tsori, Y.; Tournilhac, F.; Andelman, D.; Leibler, L.
{\it Phys. Rev. Lett.} {\bf 2003}, {\it 90}, 145504.

\bibitem{TTL} Tsori, Y.; Tournilhac, F.; Leibler, L. Submitted to {\it
Macromolecules} (2003).

\bibitem{russell3} Morariu, M. D.; Voicu, N. E.; Sch\"{a}ffer, E.;
Lin, Z.; Russell, T. P.; Steiner, U. {\it Nature Materials} {\bf
2003}, {\it 2}, 48.

\bibitem{hashimoto} Sivaniah, E.; Hayashi, Y.; Iino, M.; Fukunaga, K.;
Hashimoto, T. Submitted to {\it Macromolecules}, (2002).

\bibitem{TJ1} Turner, M. S.; Maloum, M.; Ausserr\'{e}, D.;
Joanny, J. -F.; Kunz, M. {\it J. Phys. II France} {\bf 1994}, {\it
4}, 689.

\bibitem{collin} Martinoty, P.; Gallani, J. L.; Collin, D. {\it
Phys. Rev. Lett.} {\bf 1998}, {\it 81}, 144.

\bibitem{ligoure} Bougket, G.; Ligoure, C. {\it Eur. Phys. J. B}
{\bf 1999}, {\it 9}, 137.

\bibitem{TJ2} Turner, M. S.;  Joanny, J.-F. {\it Macromolecules}
{\bf 1992}, {\it 25}, 6681.

\bibitem{turner3} Li, Z.; Qu, S.; Rafailovich, M. H.; Sokolov, J.;
Tolan, M.; Turner, M. S.; Wang, J.; Schwarz, S. A.; Lorenz, H.;
Kotthaus, J. P.  {\it Macromolecules} {\bf 1997}, {\it 30}, 8410.

\bibitem{chakra1} Podariu, L.; Chakrabarti, A. {\it J. Chem.
Phys.} {\bf 2000}, {\it 113}, 6423.
%
%
%
%
%
%
%
%
%
%
%
%
%
%
%
%
%
%
%



\end{thebibliography}
\end{document}